\documentstyle [12pt,aaspp4]{article}    

\begin{document}

\title{A Wide-Field CCD Survey For Centaurs and Kuiper Belt Objects}  
\author{Scott S. Sheppard,  David C. Jewitt, Chadwick A. Trujillo}    
\affil{Institute for Astronomy, University of Hawaii, \\
2680 Woodlawn Drive, Honolulu, HI 96822 \\ sheppard@ifa.hawaii.edu, jewitt@ifa.hawaii.edu, chad@ifa.hawaii.edu}

\author{Michael J. I. Brown}
\affil{School of Physics, University of Melbourne, \\
Parkville, Vic. 3010, Australia \\ mbrown@isis.ph.unimelb.edu.au}

\and

\author{Michael C. B. Ashley}
\affil{School of Physics, University of New South Wales, \\ 
NSW 2052, Australia \\ mcba@newt.phys.unsw.edu.au}


\begin{abstract}  

  A modified Baker-Nunn camera was used to conduct a wide-field survey
of 1428 deg$^{2}$ of sky near the ecliptic in search of bright Kuiper
Belt objects and Centaurs.  This area is an order of magnitude larger
than any previously published CCD survey for Centaurs and Kuiper Belt
Objects.  No new objects brighter than red magnitude $m_{R}=18.8$ and
moving at a rate $1\arcsec$~hr$^{-1}$ to $20\arcsec$~hr$^{-1}$ were
discovered, although one previously discovered Centaur 1997~CU26
Chariklo was serendipitously detected.  The parameters of the survey
were characterized using both visual and automated techniques.  From
this survey the empirical projected surface density of Centaurs was
found to be $\Sigma_{C}(m_{R}\leq 18.8)=7.8(^{+16.0}_{-6.6})\times
10^{-4} \mbox { deg}^{-2}$ and we found a projected surface density
$3\sigma$ upper confidence limit for Kuiper Belt objects of
$\Sigma_{K}(m_{R}\leq 18.8)<4.1\times 10^{-3} \mbox{ deg}^{-2}$.  We
discuss the current state of the cumulative luminosity functions of
both Centaurs and Kuiper Belt objects.  Through a Monte Carlo
simulation we show that the size distribution of Centaurs is
consistent with a $q\sim 4$ differential power law, similar to the
size distribution of the parent Kuiper Belt Objects.  The Centaur
population is of order $1\times 10^{7}$ (radius $\geq$1 km) assuming a
geometric albedo of 0.04.  About 100 Centaurs are larger than 50 km in
radius, of which only 4 are presently known.  The current total mass
of the Centaurs is $10^{-4}$~$M_{\oplus}$.  No dust clouds were
detected resulting from Kuiper Belt object collisions, placing a
$3\sigma$ upper limit $<600$ collisionally produced clouds of
$m_{R}<18.8$ per year.

\end{abstract}


\section{Introduction}
The outer Solar System has recently been found to be populated by a
vast number of minor bodies (e.g., Jewitt, Luu, \& Chen 1996, Jedicke
\& Herron 1997, Jewitt, Luu, \& Trujillo 1998).  As of June 2000 there
are 20 objects, known as Centaurs, which have semi-major axes and
perihelia between Jupiter and Neptune and over 300 objects, known as
Kuiper Belt objects (KBOs), which have semi-major axes greater than
that of Neptune.  Centaurs are believed to have originated from the
Kuiper Belt and are now in dynamically unstable orbits that will lead
either to ejection from the solar system, an impact with a planet or
the Sun, or evolution into a short period comet.  The brightest known
KBO, other than Pluto-Charon, has apparent red magnitude $m_{R}\sim
20$, whereas several Centaurs are brighter than this.  With current
technology, physical and chemical properties of only the biggest and
brightest minor bodies in the outer solar system can be studied.
Centaurs and KBOs could be the most primitive bodies in the solar
system and represent a unique opportunity to probe the early history
of the local solar nebula.

The Cumulative Luminosity Function (CLF) describes the sky-plane
surface density of objects brighter than a given magnitude.  Little
work has been done on the CLF of Centaurs (Jewitt et al. 1996, Levison
\& Duncan 1997), but the CLF of KBOs is well-defined between
magnitudes $20\lesssim m_{R}\lesssim26$ (Gladman~et~al. 1998, Jewitt
et al. 1998, Luu \& Jewitt 1998, Chiang \& Brown 1999).  The bright
end of the KBO CLF is very uncertain because bright KBOs are rare and
a limited amount of sky has been searched.  The extrapolated CLF
currently predicts about one KBO of $m_{R}\leq19$ per 500 square
degrees of sky.  Observations with possible bearings on large bodies
in the outer Solar System include the high eccentricities and
inclinations of KBOs which could have come from massive planetesimals
in the outer Solar System (Morbidelli \& Valsecchi 1997) and the
obliquity of Uranus's spin axis that might imply a collision with a
massive body (Singer 1975).  Evolution of the outer Solar System
suggests a more massive disk was needed in the primordial Kuiper Belt
to explain its current population size structure (Stern 1996a).  A
detailed accretion model of the early Kuiper Belt by Kenyon \& Luu
(1999) showed several Pluto sized objects could form within the age of
the Solar System.  Direct evidence for large objects in the outer
Solar System includes the large irregular satellite Triton, as well as
the Pluto-Charon system.  To constrain the CLF of both the Centaurs
and KBOs at $m_{R}\leq 20$ requires searches covering large sky areas.
To date large surveys have been conducted using photographic plates
which are difficult to parametrize.  A list of surveys which have
covered $>50$ deg$^{2}$ and which had temporal sampling sufficient to
detect Centaurs and/or KBOs is given in Table~1.  Considering that
relatively little of the sky has been searched for these Slow Moving
Objects (SMOs), and that recent advances in CCD technology have made
large area surveys feasible, this is an excellent time to search for
bright SMOs.  Accordingly, we undertook a wide-field survey in search
of bright objects in the outer Solar System.

\section{Observations}

The 0.5 meter diameter Automated Patrol Telescope (APT) at Siding
Spring Observatory in Australia was used during 3 observing runs in
early 1999.  The APT is a modified Baker-Nunn camera using a 3-element
lens to achieve a corrected wide field of view (Carter et al. 1992).
It was originally designed to track satellites in Earth orbit.  We
used a $770\times 1152$ pixel CCD manufactured by EEV with 22.5
$\micron$ pixels and a pixel scale of 9.41$\arcsec$/pixel at the f/1
focal plane for imaging.  The quantum efficiency of the detector is
0.46 in the R-band.  The detector has a $3\arcdeg \times 2\arcdeg$
rectangular field of view with the east-west direction aligned
parallel to the long axis.  Exposures were taken using a Johnson
R-band filter and the telescope was tracked at sidereal rate.  The APT
site is exceptionally dark.  The seeing was usually around
1.5$\arcsec$ as measured from other telescopes on the mountain.  At
the APT the large pixel scale did not allow us to directly measure the
seeing and moderate variations in seeing were not noticeable.  The
data were undersampled which made it difficult to distinguish between
galaxies and stars (Figure~1).

A total of 238 independent fields covering 1428 deg$^{2}$ of sky were
imaged during observing runs centered around new moon in January,
February, and March 1999 (Table~2).  A map of the sky areas searched
is shown in Figure~2 and the specific field centers are given in
Table~3 (the complete extended Table~3 is provided in the electronic
version of this paper).  The fields were selected to be observed at an
airmass $\leq 1.5$ and be within 30 degrees of the ecliptic.
Selection of the fields was made with no prior knowledge of known SMOs
in the area.  The search was made between $+10$ and $-30$ degrees of
the ecliptic.  The observational strategy was to search within $\pm
1.5$ hr of opposition.  At opposition the movement of an object is
simply related to heliocentric distance by
\begin{equation}
\frac{d\theta}{dt}\approx\frac{148}{R+R^{1/2}} \label{eq:motion} 
\end{equation}
where $R$(AU) is the heliocentric distance and $d\theta/dt$ (arcsec
hr$^{-1}$) is the rate of change of apparent position of the SMO on
the CCD.  From Equation~\ref{eq:motion}, KBOs with $R>30$~AU will have
$d\theta/dt$ $\leq 4\arcsec$~hr$^{-1}$ ($\leq 10$ pixels per day) and
Centaurs with $5<R<30$~AU have $4\arcsec$~hr$^{-1} \leq d\theta/dt
\leq 20\arcsec$~hr$^{-1}$ (between 10 and 50 pixels per day).
Therefore we sought KBOs and Centaurs by comparing images taken on
successive nights, allowing the detection of objects moving at rates
between $1\arcsec$ hr$^{-1}$ and $20\arcsec$ hr$^{-1}$.  Trailing of
both KBOs and Centaurs was negligible due to the large pixels, slow
movement of the objects, and short integration times.

Three consecutive exposures of 200 seconds each were taken on each of
3 separate nights yielding 9 images per field.  Weather permitting,
the fields were imaged on consecutive nights.  The images were bias
and dark subtracted, then flat fielded with median twilight skyflats.
The 3 images of a field taken during a single night were registered,
then combined using the IRAF ``crreject'' cosmic ray rejection
algorithm to remove cosmic rays and increase signal-to-noise.  These 3
images were taken consecutively so there was little or no telescope
motion between them.  Landolt (1992) standard stars were used for
photometric calibration.

Field centers differed by up to 50 arcseconds from night to night due
to APT pointing variations.  We aligned the fields from different
nights by eye.  The data were analyzed in two complementary ways, a)
by visual ``blinking'' and b) using an automated search algorithm.
For the visual analysis, four criteria were used for SMO
identification: (1) the object must appear in all three frames, (2) it
must have linear motion, (3) it must have a steady brightness in all 3
frames, and (4) it must have constant velocity appropriate for an SMO
($<20\arcsec$ hr$^{-1}$).  Many asteroids between magnitudes 10 and 19
were quickly identified and rejected by their large motions ($>$50
pixels in 24 hours).  They were not followed up with further
observations because of limited telescope time.  Centaur 10199
Chariklo (provisional designation 1997~CU26) was at $m_{V}=17.8$ when
detected serendipitously (Figure 1).  It was discovered independently
by the visual and automated procedures with no knowledge of its
location in the field.  No other known SMOs in the area searched were
bright enough to be detected by our survey.  It should also be noted
that the January fields were crowded due to the close proximity of the
Galactic plane to opposition, while the March fields were hampered by
bad weather.  During poor weather, the few fields with gaps of 2 or 3
days were observed on 4 nights instead of the usual 3 to ensure that
SMO identification efficiency was not significantly reduced compared
to fields taken on 3 consecutive nights.

The limiting magnitude of the survey was found by randomly placing 100
artificial SMOs onto a subset of the images.  A Moffat profile (Moffat
1969) matched to the point spread function of the APT was used to
generate the artificial SMOs.  The visual detection efficiencies for
KBOs and Centaurs are shown as a function of magnitude in Table~4.  A
KBO detection efficiency equal to half the maximum efficiency was
reached at $m_{R}=18.8$, which we take as the limiting magnitude of
this survey.  The Centaur detection efficiency equal to half the
maximum efficiency was also reached near $m_{R}=18.8$.

As a complement to the visual SMO search, a moving object detection
program was used to flag potential SMO candidates for
scrutinization. The algorithm was optimized for our undersampled
data. Objects were detected in each image by searching for maxima in
the center of a $3 \times 3 $ pixel box. An approximate
signal-to-noise and flux for the object was then determined using all
pixels within the $3 \times 3$ pixel box with flux more than $1.5
\sigma$ above the sky background. While this underestimated the flux
of extended objects, it provided a good estimate of the flux of faint
undersampled stellar images, which occupied 1 to 4 pixels depending on
the location of the image centroid with respect to the pixel
grid. Using catalogues of objects in each field with signal-to-noise
greater than 4, the software removed stationary objects by searching
for objects within $\pm 1$ pixel of the same position in multiple
images. Stationary objects with fluxes that varied by more than a
factor of 2 are retained in the catalogue as these could be blended
objects.  The program then flags ``objects'' with a consistent rate of
motion between $1^{\prime \prime}$~hr$^{-1}$ and $12^{\prime
\prime}$~hr$^{-1}$ within $20^\circ$ of the projected direction of the
ecliptic.  Candidates with fluxes varying by more than a factor of 5
were rejected as these were dominated by spurious objects and variable
stars.

The software successfully detected Centaur Chariklo, which was not
unexpected as it was more than a magnitude brighter than the survey
limit. As with searching for candidates by eye, the detection
efficiency was determined by searching for artificial SMOs. The
average detection efficiency as a function of magnitude is summarized
in Table~5. The KBO detection rate of half the maximum value was
reached at $m_{R}=18.5$ and is the limiting magnitude for the
automated detection technique.  The code is not as effective as the
eye at deblending merged objects, so the $\sim 50\%$ magnitude limit
is $\sim 0.3$ magnitudes brighter than that determined by eye.
However, in high Galactic latitude fields, the efficiency of the code
improved significantly with $\sim 70\%$ of $m_R<18.5$ KBOs being
detected.

\section{Results and Discussion}

A total of 238 fields were examined covering 1428 square degrees of
sky (about 3.5$\%$ of the whole sky).  No new SMOs were detected.  One
previously discovered Centaur (Chariklo) was serendipitously detected.
From the one detected Centaur and the efficiency of this survey we
found the empirical projected surface density of Centaurs to be
\begin{equation}
\Sigma_{C}(m_{R}\leq 18.8)=7.8(^{+16.0}_{-6.6})\times 10^{-4} \mbox{ deg}^{-2}.  \label{eq:censurf}
\end{equation}
Assuming Poisson statistics, the $3\sigma$ upper confidence limit for
the projected surface density of KBOs from our survey is
\begin{equation}
\Sigma_{K}(m_{R}\leq 18.8)<4.1\times 10^{-3} \mbox{ deg}^{-2}.  \label{eq:kbosurf}
\end{equation}
We plot these and cumulative surface density measurements from other
surveys for Centaurs and KBOs in Figures~3 and~5, respectively.  Error
bars were calculated at $1\sigma$ confidence from detection
efficiencies and Poisson error statistics when not explicitly given in
the published papers (Gehrels 1986).  Quoted upper limits are Poisson
$3\sigma$ confidence limits.  The CLF is modeled by
\begin{equation}
\mbox{log}[\Sigma (m_{R})]=\alpha (m_{R}-m_{o})  \label{eq:slope}
\end{equation}
where $\Sigma (m_{R})$ is the number of objects per square degree
brighter than $m_{R}$, $m_{o}$ is the magnitude at which $\Sigma =
1$~deg$^{-2}$, and 10$^{\alpha}$ describes the slope of the luminosity
function.

\subsection{The Centaurs}
The current state of the CLF of Centaurs is shown in Figure~3.  Almost
half of the known Centaurs have been found by the Spacewatch program
(Jedicke \& Herron 1997, McMillan 1999).  The parameters of the
Spacewatch survey have changed with time as the detector and observing
strategy have been modified.  Dr. Robert McMillan, head of the
Spacewatch program, kindly provided estimates of the Spacewatch survey
parameters (Table~1).  Spacewatch points in Figures 3 and 4 were
calculated using the total area searched with the number of Centaurs
found.  They are lower limits since the effective area searched is
less than the reported total area searched due to some observations in
marginal observing conditions and some areas covered multiple times.
Kowal's (1989) Centaur constraint is plotted with a sizeable
uncertainty because of the large trailing losses incurred in long (75
minute) photographic integrations on (relatively) fast moving
Centaurs.  The CLF of the Kuiper Belt (c.f. section 3.2) is
overplotted on the Centaur data in Figure~3.  A line of the same slope
has been vertically shifted to run through the points having one or
more Centaur detections.  The offset line is consistent with every
observational point shown in Figure~3, within the uncertainties of
measurement.  The Centaurs reach $\Sigma = 1$ deg$^{-2}$ at
$m_{R}=24.6\pm 0.3$.  Thus, Centaurs have an apparent
sky-plane-density about 10 times less than that of the KBOs at a given
magnitude.

The CLF represents the convolved effects of the size and distance
distributions of the objects.  A Monte Carlo simulation was used to
determine the total number of Centaurs larger than a certain radius.
The Centaurs were assumed to follow a differential power-law radius
distribution of the form $n(r)dr=\Gamma r^{-q}dr$, where $\Gamma$ and
$q$ are constants, $r$ is the radius of the Centaur, and $n(r)dr$ is
the number of Centaurs with radii in the range $r$ to $r+dr$.  The
radii of the Centaurs were assumed to lie between $r_{\mbox{min}}$ and
$r_{\mbox{max}}$.  To find the total number of Centaurs, we adopted
$r_{\mbox{min}}=1$~km (since the smaller short period comets are about
this size), and $r_{\mbox{max}}=$ 1000~km (about the size of the
largest known Trans-Neptunian body, Pluto).  The derived total number
of Centaurs is insensitive to $r_{\mbox{max}}$.  The surface density
distribution of Centaurs in the plane of the ecliptic varies with
heliocentric distance and was assumed to be a power law of the form
$n(R)dR =\Sigma_{o}R^{-\gamma}dR$, where $\Sigma_{o}$ and $\gamma$ are
constants, $R$ is the heliocentric distance, and $n(R)dR$ is the
number of objects with heliocentric distance in the range $R$ to
$R+dR$.  The heliocentric distance was taken between 10 and 30 AU.  A
$\gamma = -1.3$ was chosen as the most likely Centaur distance
distribution based on the simulation by Levison and Duncan (1997).

The Monte Carlo simulation was performed by randomly choosing a
distance and radius for up to $10^{8}$ Centaurs from the two power law
equations above.  Each Centaur's apparent magnitude at opposition was
then calculated using
\begin{equation}
m_{R}=m_{\odot}-2.5\mbox{log}(p_{R}r^{2}\phi (\alpha )/2.25\times 10^{16}R^{2}\Delta ^{2})   \label{eq:appmag}
\end{equation}
in which $r$ is in km, $R$ is in AU, $\Delta$ is the geocentric
distance in AU taken here to be $\Delta =R-1$, $m_{\odot}$ is the
apparent red magnitude of the sun ($-27.1$), $m_{R}$ is the apparent
red magnitude of the Centaur, $p_{R}$ is the geometric red albedo
assumed to be 0.04, and $\phi (\alpha)$ is the phase function in which
the phase angle $\alpha=0$ deg at opposition and thus $\phi
(0)=1$.

The results of the Monte Carlo simulation using $q=3.5$, 4.0, and 4.5
are shown in Figure~4.  We obtained the total number of Centaurs with
$r>1$~km by normalizing the simulation to the CLF of the Centaurs at
$m_{R}=22.0$.  We assumed the inclination distribution extends to $\pm
30$ degrees in latitude from the ecliptic: the total sky area occupied
by Centaurs is then about 20000 deg$^{2}$.  We used the relation
$N(>r)=\Gamma /(q-1)r^{q-1}$ where $N$ is the cumulative number of
Centaurs with radius greater than $r$ to find $\Gamma$.  Table~6 shows
the results for the number of Centaurs using different assumed values
of the main parameters.  For the $q=4$ index we found there are about
$1\times 10^{7}$ Centaurs with $r>1$~km.  Uncertainties in $q$ of $\pm
0.5$ correspond to uncertainties in the population of factors 3 to 4.
Plausible uncertainties in $\gamma$ have a smaller effect (Table~6).

The $q=3.5$, 4.0 and 4.5 curves are all consistent with the Centaur
CLF (Figure~4).  If we use the closest fit $q=4$ simulation we obtain
$N(r>50$~km$)\sim 100$ (Table~6).  The number of Centaurs with
$r>50$~km is comparable to the 300 main-belt asteroids (Tedesco~1989)
in this size regime but much smaller than the $7\times 10^{4}$ KBOs
expected with $r>50$~km (Jewitt et al. 1998).  Our result is about an
order of magnitude smaller than the result obtained by Jewitt et
al. (1996).  These authors detected 2 Centaurs in a narrow field
survey and crudely estimated the Centaur population based on only 1 of
the Centaurs and in the absence of a model of the Centaur radial
distance distribution.  Jedicke \& Herron (1997) found, at the $99\%$
confidence level, that there must be fewer than 250 Centaurs with
$r>50$~km (we scaled the quoted value for $r>25$~km in Jedicke \&
Herron (1997) to $r>50$~km using a q=4 index for the size
distribution) and they also concluded there are 3 or fewer Centaurs
with $r>100$~km.  Although all three results are consistent, our
simulation using the CLF in Figure~3 presumably provides a superior
estimate of the population since more Centaurs are now known and a
model of the distance distribution has been incorporated.

Assuming the Centaurs are supplied by the KBOs, in steady-state we
should have $N_{c}\sim N_{k}t_{c}/t_{k}$ where $N_{c}$ is the number
of Centaurs, $N_{k}$ is the number of KBOs, $t_{c}$ is the average
Centaur lifetime, and $t_{k}$ is the average KBO lifetime.  For KBOs
$t_{k}\approx 10^{10}$ years (Quinn et al. 1990) and for Centaurs
$t_{c}\approx 10^{6}$ to $10^{7}$ years (Hahn \& Bailey 1990, Asher \&
Steel 1993, Levison \& Duncan 1997).  With $N_{k}\simeq 70000$
($r>50$~km) KBOs (Jewitt et al. 1998), we expect $N_{c}\simeq 7$ to
$70$ ($r>50$~km) Centaurs based on the steady state assumption.  This
is consistent with $N_{c}\sim 100$ estimated here from the CLF, and
consistent with the Kuiper Belt supplying the Centaur population.
Since the average Centaur lifetime expressed in years is similar to
the number of Centaurs with radii greater than 1~km ($N(r>1
\mbox{km})/t_{c}\sim 1$), about one Centaur with $r\geq 1$~km must be
delivered to the gas giant region per year.  Duncan et al. (1995)
found the fraction of KBOs which leave the Kuiper Belt from planetary
interactions is $5\times 10^{-11}$ per year.  Assuming that half of
the KBOs which leave the Kuiper Belt end up as Centaurs, then with
$7\times 10^{4}$ KBOs of $r>$50 km and a Centaur lifetime of $10^{6}$
to $10^{7}$ years we would expect between 2 to 20 Centaurs to be
present with radii $>$~50~km, consistent with our results.

For a $q=4$ size distribution the total mass is
\begin{equation}
M=\frac{4\times 10^{9} \pi \rho \Gamma}{3}\left[ \frac{0.04}{p_{R}}\right] ^{3/2} \mbox{ln}\left( \frac{r_{\mbox{max}}}{r_{\mbox{min}}}\right)   \label{eq:mass}
\end{equation}
where $\rho$ (kg m$^{-3}$) is the bulk density.  Using
$\rho=1000$~kg~m$^{-3}$ (similar to cometary nuclei), $p_{R}=0.04$,
$r_{\mbox{min}}=1$~km, and $r_{\mbox{max}}=200$~km (about the size of
the largest known Centaur), we obtain $M\approx 8\times 10^{20}$ kg
($\sim 10^{-4}$~M$_{\oplus}$) for the current total mass of the
Centaurs.  Assuming Centaurs have a lifetime of $5\times 10^{6}$ years
and are in steady state we find that 0.1 M$_{\oplus}$ has been cycled
through the Centaurs during the $4.5\times 10^{9}$ year age of the
solar system.  If objects as large as Pluto ($r_{max}=1200$~km) with
$\rho=2000$~kg~m$^{-3}$ are considered then the mass cycled through
the Centaurs during the age of the solar system is about 2.5 times
greater than the above result.  Either way it is comparable to the
inferred total mass currently in the Kuiper Belt of
$\sim$0.2~M$_{\oplus}$ (Jewitt et al. 1998).

Centaurs are close enough to the Sun to show cometary activity (e.g.,
Chiron), which can lead to an overestimate of their size if assuming a
constant albedo, as we have done.  Observationally, most surveys are
only sensitive to bright (large) Centaurs.  Thus the smaller Centaur
population was obtained by extrapolating from larger Centaurs which
may not follow the same distribution.  No survey has been specifically
targeted at detecting Centaurs.  Most surveys which have found
Centaurs are optimized instead for finding KBOs or asteroids.  This
makes it hard to assess the sensitivity to Centaurs of some published
surveys.  Surveys plotted in Figure~3 are those which were capable of
finding Centaurs in at least the range of heliocentric distances 10~AU
to 30~AU.  Our survey was sensitive to Centaurs from 5 to 30~AU.  Only
4 of the 20 currently known Centaurs could have radii $>50$~km
(assuming geometric albedo 0.04) and our simulation predicts $\sim
100$ such objects in the whole sky. Most of them would be located near
but inside the orbit of Neptune, with apparent red magnitudes near 23.
They would have naturally escaped detection in most wide field surveys
conducted to date.

\subsection{The Kuiper Belt}
Surface densities of KBOs from different surveys are in reasonable
agreement between $20\lesssim m_{R}\lesssim26$ (Figure~5).  A weighted
least squares fit to the CLF for all the data with at least one
detection is shown in Figure~5 (we used only the middle of the three
Tombaugh points in the fit), yielding $\alpha = 0.59\pm 0.05$,
$m_{o}=23.0\pm 0.2$.  The fit is consistent with the fits found by
Jewitt~et~al. (1998) [$\alpha\simeq 0.55\pm 0.05$, $m_{o}\simeq
23.25\pm 0.11$], Gladman~et~al. (1998) [$\alpha=0.76^{+0.1}_{-0.11}$,
$m_{o}=23.40^{+0.20}_{-0.18}$], and Chiang \& Brown (1999)
[$\alpha=0.52\pm 0.02$, $m_{o}=23.5\pm 0.06$].  Our survey's $3\sigma$
upper limit is consistent with the constant slope of the CLF but does
not rule out a steepening in the CLF for $m_{R}<19$.

The above and some other previous works have improperly calculated the
surface density from Tombaugh's (1961) survey (see Table~1 for the
correct results of Tombaugh's survey).  They used Tombaugh's follow up
survey which was conducted to search for satellites or other objects
that might be near Pluto.  This survey reached $m_{R}=16.8$ and
covered 1530 square degrees of sky.  However, Pluto was discovered in
a survey which reached only $m_{R}=15.5$ and covered about 28000
deg$^{2}$, initiated earlier and continued later (Table 1).  The 1530
square degree survey should be interpreted as a null result since
Pluto was not detected serendipitously.  Therefore, the Tombaugh datum
used in some earlier works should be shifted to brighter magnitudes
and lower surface densities.

\subsection{Collision Clouds}
The Kuiper Belt was first thought to be collisionless though it now
appears collisional effects might be very important (Davis \&
Farinella 1997).  This has the result of removing the primitive and
possibly unprocessed material from KBOs and might allow for dust to be
observed in the Kuiper Belt.  The collision products might be seen
directly by the detection of dust clouds around the KBOs (Alcock \&
Hut 1993, Stern 1996b).  Assuming there are $10^{9}$ KBOs with
$r>1$~km and that the Kuiper Belt extends $\pm30$ degrees from the
ecliptic, Stern (1996b) and Alcock \& Hut (1993) predicted between 1
and 100 collisions per year producing detectable ($m_{R}\lesssim$ 15)
dust clouds.

The characteristic expansion time of a cloud is $t\approx r/v$, where
$v$ is the expansion velocity and $r$ is the radius of the expanding
cloud.  The velocity of the ejected material depends on the energy of
the impact and size of the projectile and target.  KBOs have Keplerian
velocities of a few kilometers per second and escape velocities for
1~km objects are around 1~m~s$^{-1}$.  The cloud should reach peak
brightness in reflected light when its optical depth is unity.  The
time to reach peak brightness is $t\approx c/v$, where $c$ is the
critical radius at which optical depth unity is reached.  The critical
radius depends on the size distribution of ejected particles.  If
constant density is assumed then the critical radius is just the
square root of the geometric cross-section of the particles
($c=a[a_{i}/a]^{1.5}$ where $a$ is the average radius of particles
created from the initial projectile radius of $a_{i}$).  The duration
of visibility depends on the limiting magnitude.  Using typical values
of $v=1$ to 5~km~s$^{-1}$ (the approximate expansion velocity of gas
produced by impact vaporization) and a constant size of 0.1 to
10$\micron$ for particles in the cloud we find that a dust cloud
should reach optical depth unity in only a few days and be brighter
than red magnitude 19 for about 20 to 100 days.  If 100 clouds are
produced in one year and each is visible for 50 days about 14 clouds
should be visible at any instant.  With 14 clouds in 20,000 square
degrees, we expect to find one cloud for every 1430 square degrees of
sky.  Owing to the large pixels of the APT our survey would not
resolve one of these collision clouds if in our field of view, though
we would detect the KBO type movement of any clouds brighter than
$m_{R}=18.8$.  No dust clouds brighter than $m_{R}=18.8$ were detected
in 1428 square degrees of sky searched in this survey.  Assuming
Poisson spatial distribution this gives a $3\sigma$ confidence limit
that there are $<82$ clouds for 20,000 square degrees at any given
time or less than 1 cloud per 244 square degrees of sky with
$m_{R}<18.8$.  This also puts a $3\sigma$ limit of $<600$ clouds
produced per year with $m_{R}<18.8$.

More sky must be searched to determine if Pluto is just one of several
other large KBOs waiting to be discovered.  If the amount of sky
searched for slow moving objects were doubled, strong constraints on
the bright end of the KBO CLF would be obtained.  This survey was
hindered by the large pixel size of the CCD.  Currently CCDs four
times as large and with pixels half the size are routinely available.
It is our goal to continue to survey the sky for bright SMOs with a
bigger CCD and smaller pixels.  This will improve the efficiency of
the survey and allow the data to be better sampled.

\section{Summary}

1. A total of 1428 square degrees of sky were searched for KBOs and
Centaurs to a limiting magnitude of $m_{R}=18.8$.  One previously
discovered Centaur was detected serendipitously (1997~CU26 Chariklo),
and no KBOs were detected.

2. This survey, the largest published CCD survey to date sensitive to
Centaurs and Kuiper Belt Objects, was analyzed both by an automated
technique and by eye.  The survey parameters were well established.

3. The cumulative surface density is
$\Sigma_{C}=7.8(^{+16.0}_{-6.6})\times 10^{-4} \mbox{ deg}^{-2}$ for
Centaurs at $m_{R}=18.8$ with error bars at the $1\sigma$ level.  The
Centaur cumulative luminosity function (CLF) has a slope ($\alpha\sim
0.6$) similar to the Kuiper Belt CLF but shifted about 1.5 magnitudes
fainter.  The number of Centaurs per unit sky area is about 1/10th the
number of KBOs, at any given magnitude.

4. A Monte Carlo simulation was performed to obtain the expected
number distribution of Centaurs.  Differential power-law size
distribution indices between q=3.5 and 4.5 are consistent with the
Centaur CLF.  Using q=4, we find there are about $1\times 10^{7}$
Centaurs with radii larger than 1 km and $\sim 100$ larger than 50 km
in radius, of which only 4 (Chiron, Pholus, Chariklo(1997 CU26), and
1995DW2) are presently known.  The combined mass of all Centaurs is
$10^{-4}$~$M_{\oplus}$.  Assuming a steady state this implies that
0.1~$M_{\oplus}$ has been cycled from the Kuiper Belt through the
Centaurs during the age of the solar system.

5. The surface density of Kuiper Belt objects at $m_{R}=18.8$ is
$\Sigma_{K}<4.1\times 10^{-3} \mbox{ deg}^{-2}$ at the $3\sigma$
confidence level.  This result is consistent with the previously
determined CLF for Kuiper Belt objects but does not rule out a
steepening in the slope for $m_{R}<19$.

6. No collisionally produced dust clouds were observed in the Kuiper
Belt.  We place a $3\sigma$ upper limit of $<600$ visible collisional
clouds ($m_{R}<18.8$) produced per year.

\section*{Acknowledgments}

  We thank Jill Rathborne for observational assistance and Robert
McMillan, Jeff Larsen, Bruce Koehn, and Steve Larson for very helpful
information about their Near Earth Object surveys.  This work was
supported by a grant to D.J. from NASA.

\newpage

\begin{figure}
\caption{Portions of three APT fields collected during the run in
February 1999, with a 1 day timebase between images.  The width of
each image section is 1/4 degree and the height is 1/8 degree.  Each
image section shows 0.5\% of the total area of one field.  The arrows
show the position of the Centaur 1997~CU26 Chariklo on each night.}
\end{figure}

\begin{figure}
\caption{Sky covered during this survey.  Shaded areas represent the
month in which the sky was searched in 1999.  Horizontal axis is Right
Ascension in hours and vertical axis is Declination in degrees.}
\end{figure}

\begin{figure}
\caption{The Cumulative Luminosity Function (CLF) of the Centaurs.
Symbols with down arrows represent 3$\sigma$ upper limits from surveys
with null results.  Symbols with up arrows are lower limits.  Other
symbols are from published surveys.  The dashed line is the fit to the
KBO data in Figure 5.  The solid line is the same fit shifted
vertically to fit the Centaur CLF.}
\end{figure}

\begin{figure}
\caption{Simulation of the Cumulative Luminosity Function (CLF) of the
Centaurs.  This simulation used parameters q=3.5, 4.0, and 4.5 with
$r_{\mbox{min}}=1$~km, $r_{\mbox{max}}=1000$~km, $\gamma = -1.3$,
$R_{\mbox{min}}=10$~AU, and $R_{\mbox{max}}=30$~AU.  The solid line is
the fit to the KBO data shifted vertically to fit the lower Centaur
CLF.  See the text for an explanation of the parameters.}
\end{figure}

\begin{figure}
\caption{The Cumulative Luminosity Function (CLF) of the Kuiper Belt.
Symbols with arrows are 3$\sigma$ upper limits from surveys with null
results.  Other symbols are from various published surveys with
1$\sigma$ error bars. The solid line is the best fit to the data.}
\end{figure}

\begin{center}
\begin{deluxetable}{lccccccc}
\tablenum{1}
\tablecolumns{8}
\tablecaption{List of large and medium sized searches for Slow Moving
Objects.\tablenotemark{a} }
\tablehead{
\colhead{Survey} & \multicolumn{2}{c}{Square Degrees\tablenotemark{b}} &
\colhead{Limiting Mag.} &
\multicolumn{2}{c}{Detected Objects\tablenotemark{c}} &
\colhead{Object(s)\tablenotemark{d}} & \colhead{Year(s)} \\
\colhead{} & \colhead{KBOs} & \colhead{Centaurs} & \colhead{($m_{R}$)} &
\colhead{KBOs} & \colhead{Centaurs} & \colhead{($m_{R}$)} & \colhead{}}
\startdata
Tombaugh(1961)\tablenotemark{e} & 1530 & 1530 & 16.8 & 0 & 0 & NA & 1939--1940 \nl
 & $\sim19500$ & $\sim19500$ & 15.5 & 1  & 0 & 14 & 1929--1945 \nl
 & $\sim25500$ & $\sim25500$ & 15 &        & 0 &  & \nl
 & $\sim28000$ & $\sim28000$ & 13 &        & 0 &  & \nl
Kowal(1989)\tablenotemark{e} & 6400 & 6400 & 19.5 & 0 & 1 & 18 & 1976--1985 \nl
This Work & 1428 & 1428 & 18.8 & 0 & 0[1] & NA & 1999 \nl
Luu et al.(1988)\tablenotemark{e} & 297 & 297 & 19.5\tablenotemark{f} & 0 & 0 & NA & 1987 \nl
Jewitt et al.(1998) & 51.5 & 51.5 & 22.5 & 13 & 0 & 20.60 & 1996--1997 \nl
Irwin et al.(1995)\tablenotemark{e} & 50 & 50 & 20 & 0 & 0 & NA & 1995 \nl
Spacewatch\tablenotemark{1} & $\sim$1000 & $\sim$10000 & 20-21\tablenotemark{f} & 7 & 8[1] & 16.5,17.4, & 1989--\nl
  & & & & & &19.5,20.4\tablenotemark{f} &  \nl
LONEOS\tablenotemark{2} & 0 & ?\tablenotemark{g} & $\sim$19 & NA & 0[2] & NA & 1998-- \nl
Catalina\tablenotemark{3} & 0 & ?\tablenotemark{h} & $\sim$19 & NA & 1[1] & 18 & 1999-- \nl
\enddata
\tablenotetext{a}{($>50$ deg$^2$)}
\tablenotetext{b}{Area covered in which KBOs ($d\theta /dt<4\arcsec$ hr$^{-1}$) and Centaurs ($4\arcsec $ hr$^{-1} < d\theta /dt <20\arcsec$ hr$^{-1}$) could be detected.}
\tablenotetext{c}{Shows number of new objects detected as well as serendipitously detected objects in brackets [].}
\tablenotetext{d}{The magnitude of the brightest object(s) detected in the survey.}
\tablenotetext{e}{Surveys were done using photographic plates.}
\tablenotetext{f}{Magnitude was calculated from quoted V-band value using $V-R=0.5$.}
\tablenotetext{g}{Covers most of sky north of $-30$ degrees declination}
\tablenotetext{h}{Covers about 300 deg$^{2}$ per good night}
\tablecomments{Spacewatch, LONEOS, and Catalina are Near Earth Object searches capable of detecting some SMOs.}

\tablerefs{(1)McMillan (1999); (2)Koehn (1999); (3)Larson (1999).}
\end{deluxetable}
\end{center}

\begin{center}
\begin{deluxetable}{ccccc}
\tablenum{2}
\tablewidth{5 in}

\tablecaption{APT survey data}

\tablecolumns{5}
\tablehead{
\colhead{Date} & \colhead{\# Nights} &
\colhead{\# Images} & \colhead{\# Fields} & \colhead{Sky Area}
\\ \colhead{} & \colhead{} & \colhead{} & \colhead{} &\colhead{(deg$^{2}$)}}
\startdata
January 1999 & 10 & 201 & 58 & 348 \nl
February 1999 & 15 & 468 & 112 & 672 \nl
March 1999 & 15 & 399 & 68 &  408 \nl
Total & 40 & 1068 & 238 & 1428 \nl
\enddata
\end{deluxetable}
\end{center}

\begin{center}
\begin{deluxetable}{ccrl}
\tablenum{3}
\tablewidth{6 in}
\tablecaption{Observations from the Automated Patrol Telescope}
\tablecolumns{4}
\tablehead{
\colhead{Field} & \colhead{RA(2000)} & \colhead{
Dec(2000)}  & \colhead{UT Date} }
\startdata
f0991 & $07:29:24$ & $ 16:00:00$ & 1999/Jan/13,14,15 \nl
f0044 & $08:35:60$ & $ 00:00:00$ & 1999/Jan/15,17,18 \nl
f3926 & $08:53:11$ & $-08:00:00$ & 1999/Feb/17,19,21 \nl
f0409 & $09:27:06$ & $ 06:00:00$ & 1999/Feb/14,15,16 \nl
f1003 & $09:59:12$ & $ 16:00:00$ & 1999/Feb/16,17,18 \nl
f4288 & $10:18:22$ & $-14:00:00$ & 1999/Feb/13,14,15 \nl
f3933 & $10:18:00$ & $-08:00:00$ & 1999/Mar/11,12,14 \nl
f3703 & $12:13:47$ & $-04:00:00$ & 1999/Mar/20,22,23 \nl
\enddata
\tablecomments{Centaur Chariklo was found in field f0409.}
\tablecomments{The complete version of this table will be in the electronic edition of the Astronomical Journal.  The printed edition contains only a sample.}
\end{deluxetable}
\end{center}

\begin{center}
\begin{deluxetable}{ccccc}
\tablenum{4}
\tablewidth{5 in}

\tablecaption{Eye detection efficiency versus magnitude for artificial
KBOs and Centaurs.}

\tablecolumns{3}
\tablehead{
\colhead{Magnitude} & \colhead{Detection Efficiency(\%)} &
\colhead{Detection Efficiency(\%)}
\\ \colhead{($m_{R}$)} &
\colhead{KBOs\tablenotemark{a}} & \colhead{Centaurs\tablenotemark{b}}}
\startdata
18.00 & 92 & 90 \nl
18.25 & 90 & 86 \nl
18.50 & 77 & 72 \nl
18.75 & 55 & 47 \nl
19.00 & 16 & 7 \nl
\enddata
\tablenotetext{a}{Artificial KBOs moved between 8 and 10 pixels per day ($3$ to $4\arcsec$ hr$^{-1}$).}
\tablenotetext{b}{Artificial Centaurs moved between 15 and 30 pixels per day ($6$ to $12\arcsec$ hr$^{-1}$).}
\end{deluxetable}
\end{center}

\begin{center}
\begin{deluxetable}{ccccc}
\tablenum{5}
\tablewidth{5 in}

\tablecaption{Automated detection efficiency versus magnitude for artificial
KBOs and Centaurs.}

\tablecolumns{2}
\tablehead{
\colhead{Magnitude} & \colhead{Detection Efficiency(\%)}
\\ \colhead{($m_{R}$)} & \colhead{SMO\tablenotemark{a}}}
\startdata
17.00 & 74 \nl
17.25 & 68 \nl
17.50 & 67 \nl
17.75 & 58 \nl
18.00 & 54 \nl
18.25 & 44 \nl
18.50 & 41 \nl
18.75 & 21 \nl
19.00 & 6 \nl
\enddata
\tablenotetext{a}{Artificial Slow Moving Objects moved between $3$ and $8\arcsec$ hr$^{-1}$.}
\end{deluxetable}
\end{center}

\begin{center}
\begin{deluxetable}{ccccccc}
\tablenum{6}
\tablewidth{6 in}
\tablecaption{Centaur Monte Carlo Simulation Results}
\tablecolumns{7}
\tablehead{
\colhead{$q$\tablenotemark{a}} & \colhead{$\gamma$\tablenotemark{b}} & \colhead{albedo}  & \colhead{$\Gamma$} & 
\colhead{N($r>1$)\tablenotemark{c}} & \colhead{N($r>50$)\tablenotemark{c}} & \colhead{N($r>100$)\tablenotemark{c}} }
\startdata
3.0 & $-1.3$ & 0.04 & $1.3\times 10^{6}$ & $6.7\times 10^{5}$ & 270 & 65 \nl
3.5 & $-1.3$ & 0.04 & $7.5\times 10^{6}$ & $3.0\times 10^{6}$ & 170 & 30 \nl
4.0 & $-1.3$ & 0.04 & $3.6\times 10^{7}$ & $1.2\times 10^{7}$ & 100 & 12 \nl
4.5 & $-1.3$ & 0.04 & $1.8\times 10^{8}$ & $5.0\times 10^{7}$ & 60 & 5 \nl
4.0 & $-0.3$ & 0.04 & $2.3\times 10^{7}$ & $7.6\times 10^{6}$ & 60 & 8 \nl
4.0 & $-0.8$ & 0.04 & $3.0\times 10^{7}$ & $1.0\times 10^{7}$ & 80 & 10 \nl
4.0 & $-1.8$ & 0.04 & $4.8\times 10^{7}$ & $1.6\times 10^{7}$ & 130 & 15 \nl
4.0 & $-2.3$ & 0.04 & $6.3\times 10^{7}$ & $2.1\times 10^{7}$ & 170 & 20 \nl
4.0 & $-1.3$ & 0.10 & $9.6\times 10^{6}$ & $3.2\times 10^{6}$ & 25 & 3 \nl
4.0 & $-1.3$ & 0.40 & $1.2\times 10^{6}$ & $3.9\times 10^{5}$ & 3 & 0 \nl

\enddata
\tablenotetext{a}{Index of power law size distribution}
\tablenotetext{b}{Index of power law distance distribution}
\tablenotetext{c}{Number of objects with radius greater than $r$ in km}
\end{deluxetable}
\end{center}

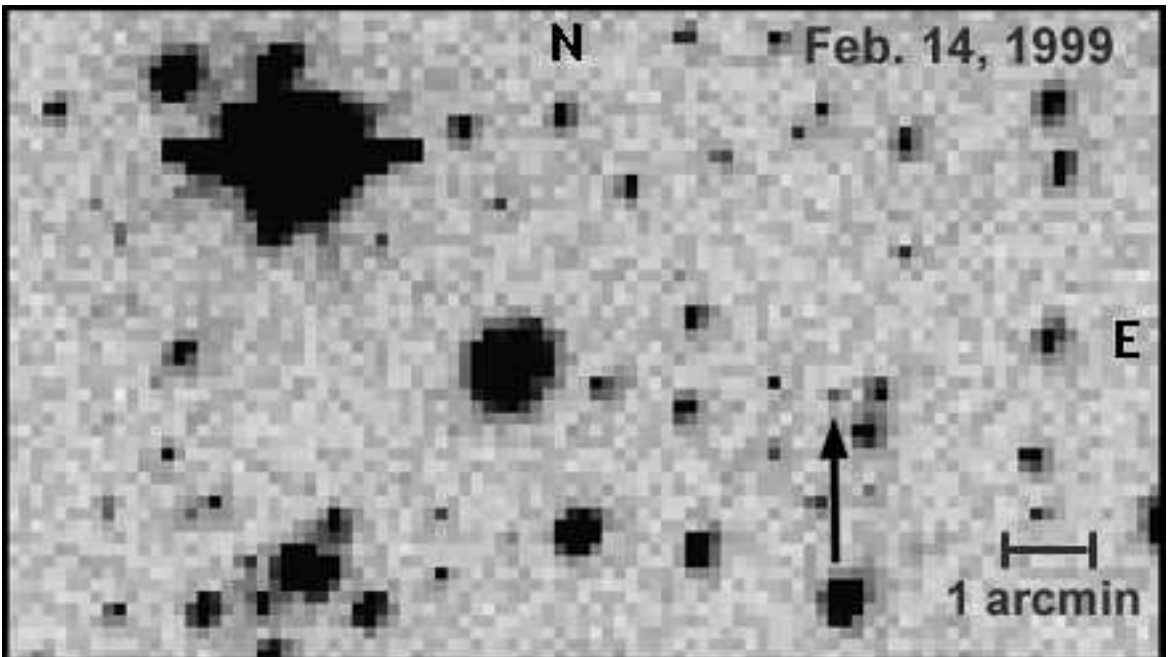
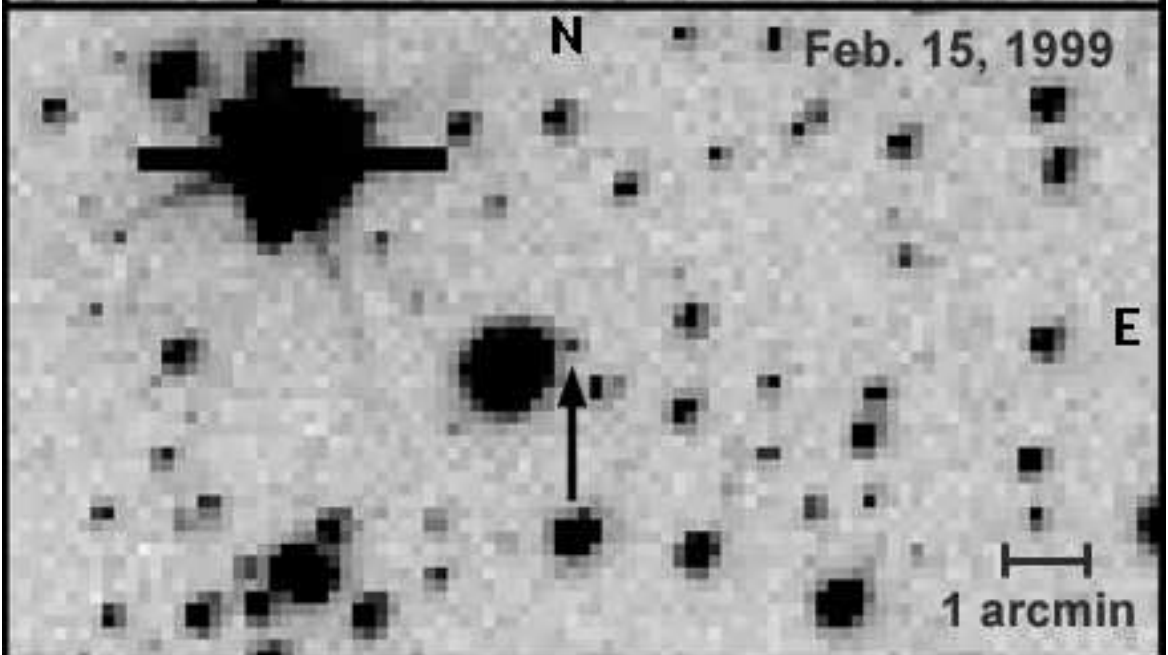
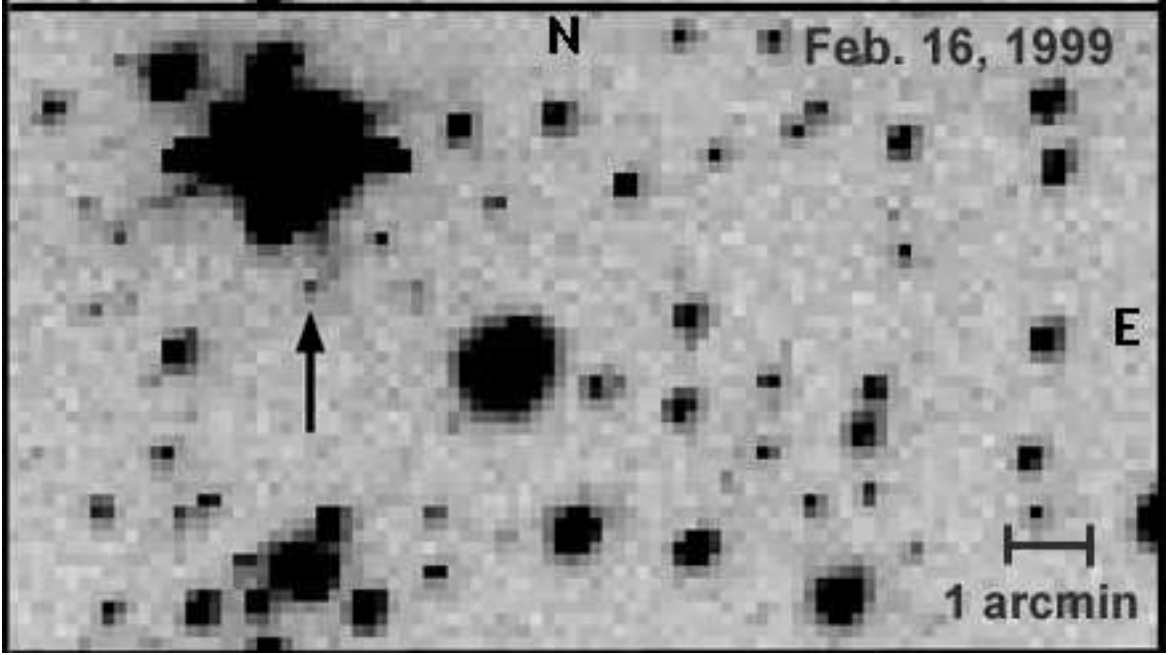
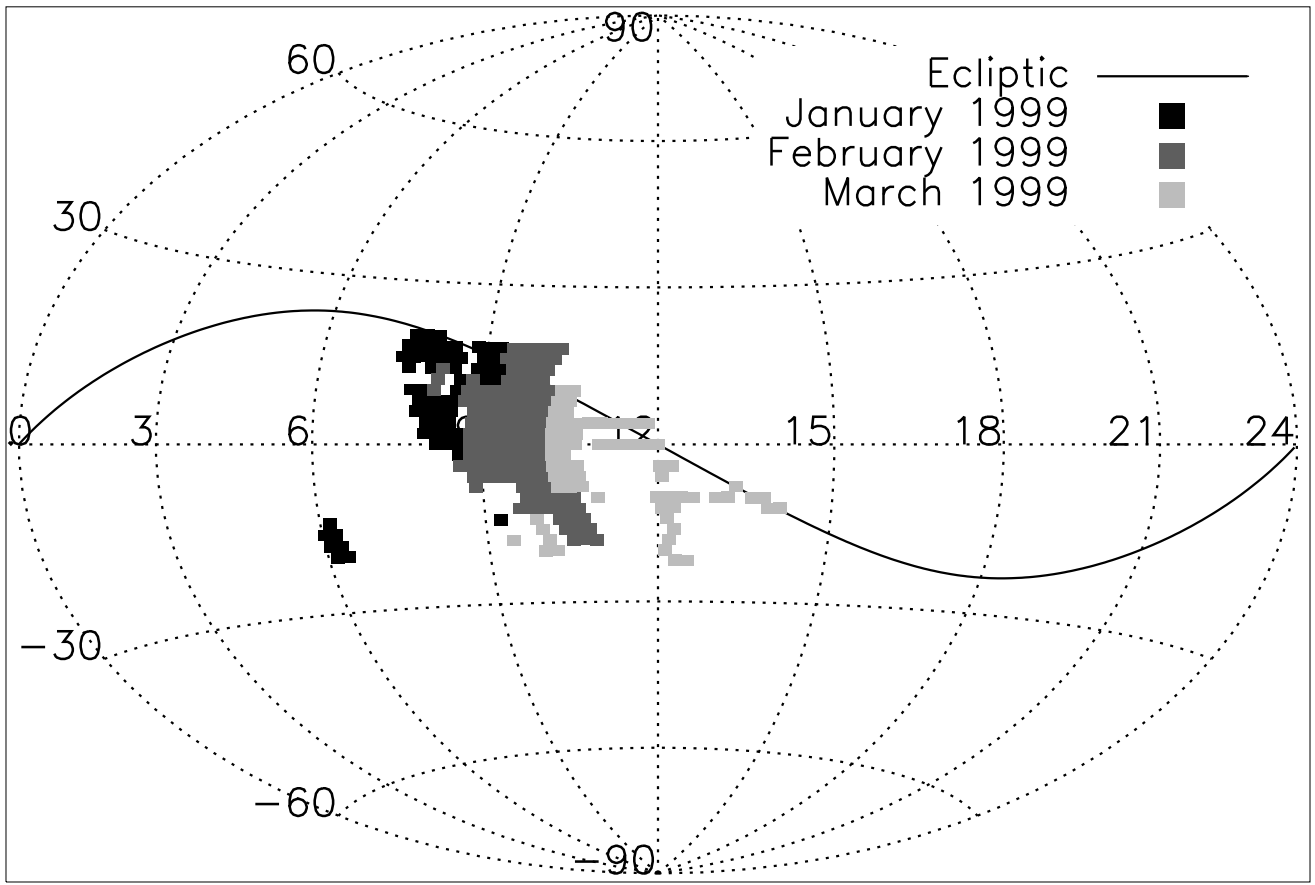
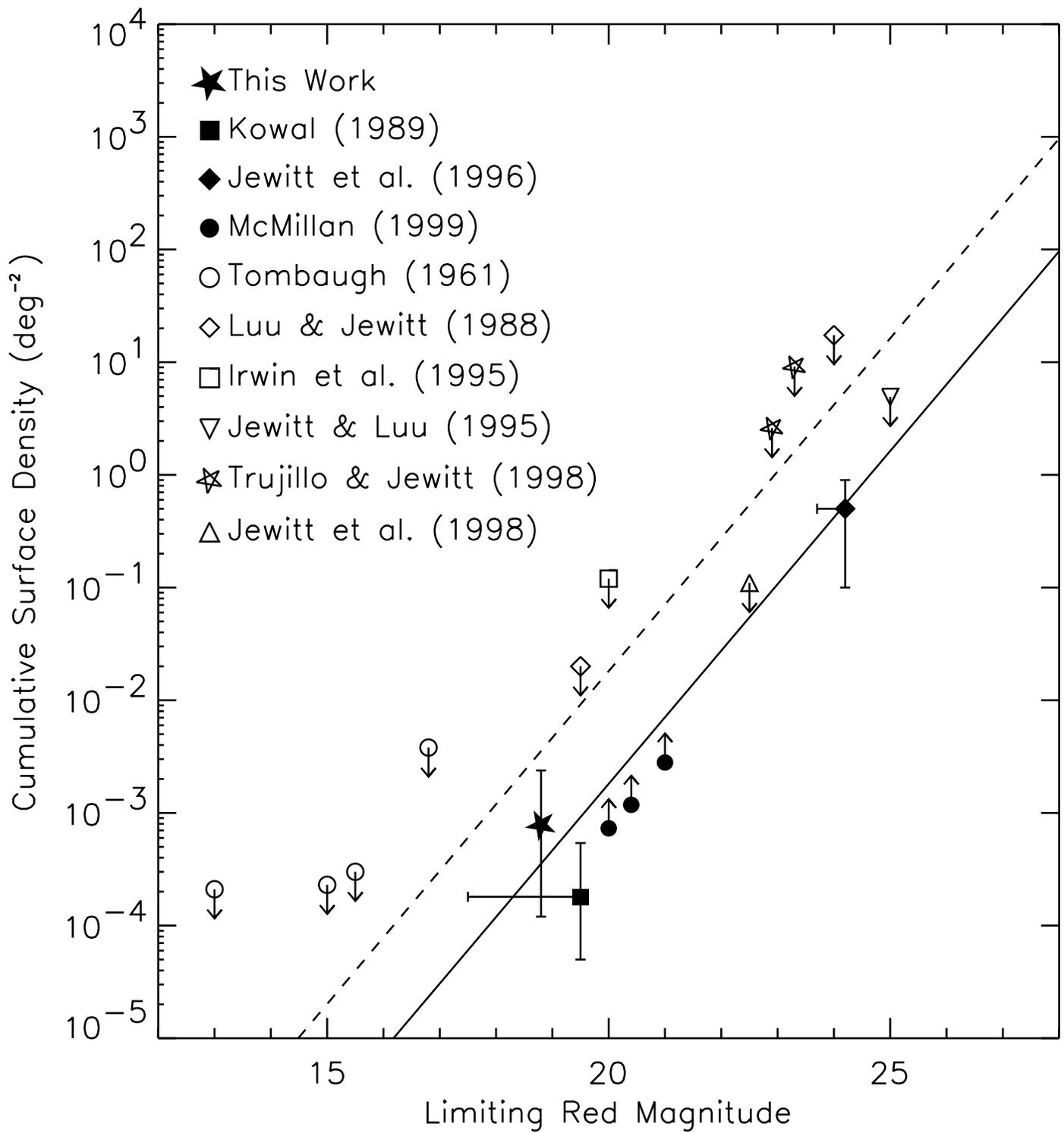
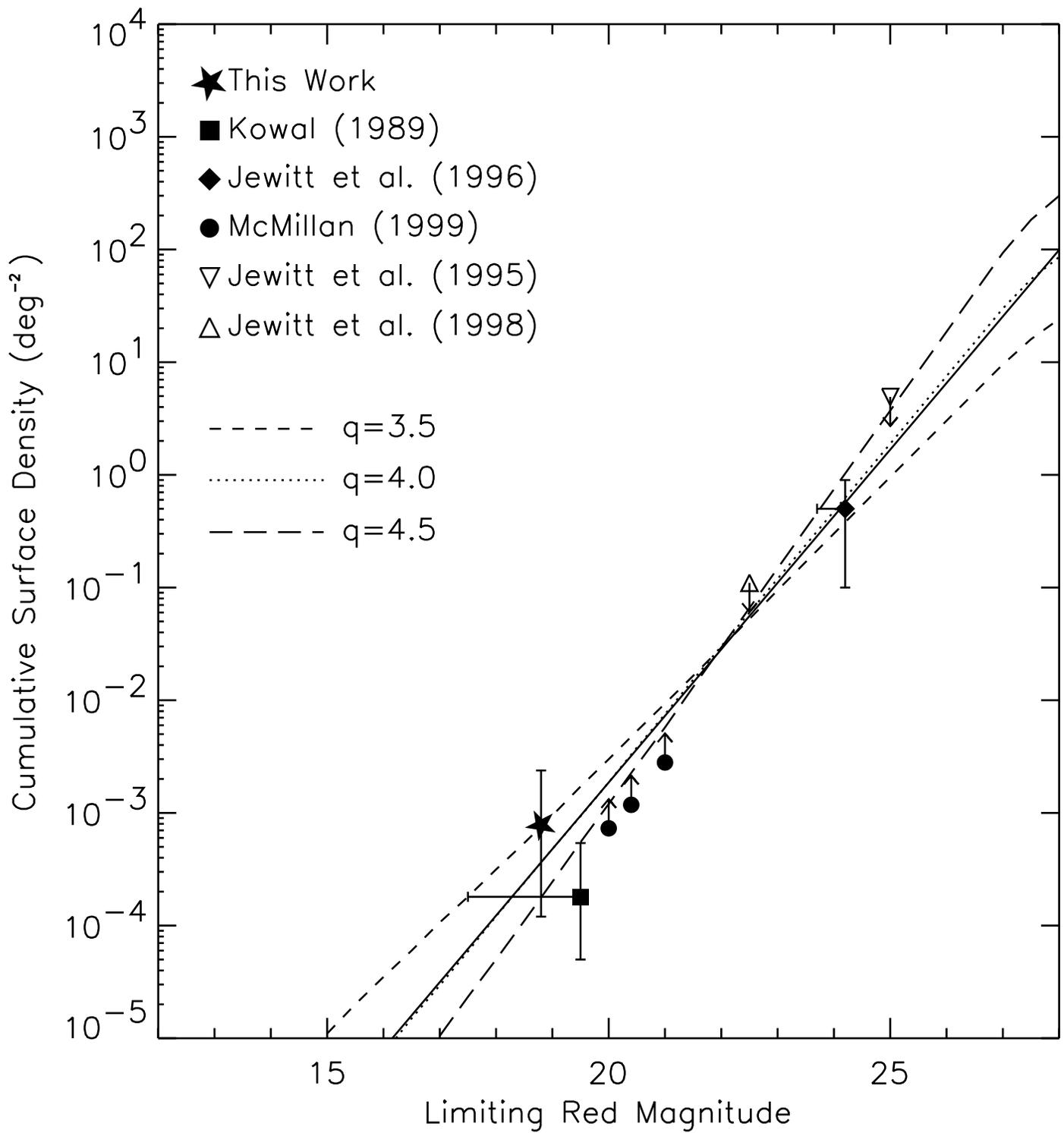
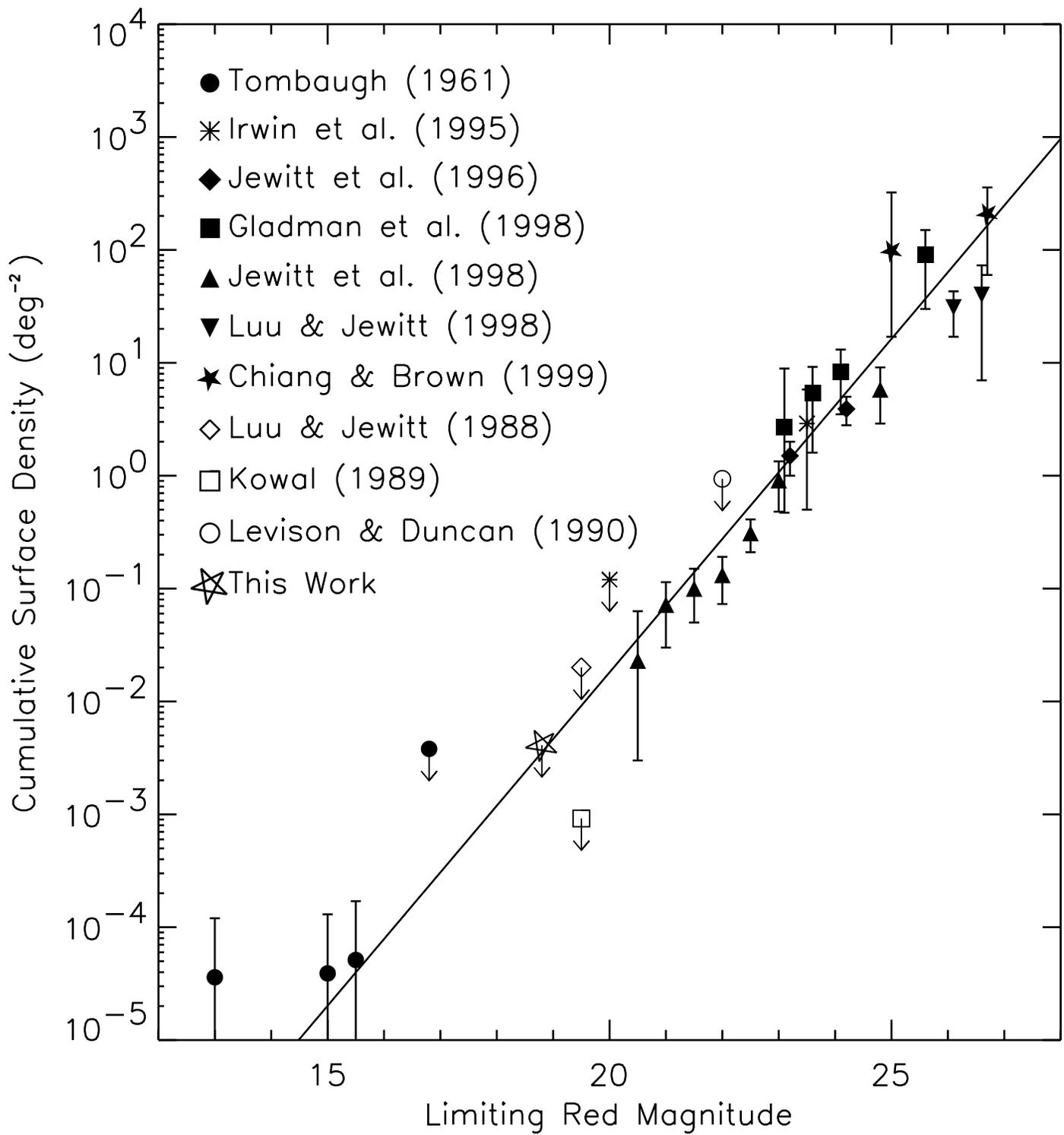
\end{document}